\documentclass{llncs}
\usepackage{graphicx}
\usepackage{float}
\usepackage{cite}
\usepackage{url}
\usepackage{soul}
\usepackage[caption=false]{subfig}

\begin{document}
	\title{VirtuWind -- An SDN- and NFV-based Architecture for Softwarized Industrial Networks}
	\titlerunning{Hamiltonian Mechanics}  
	%
	\author{
		Ermin Sakic\inst{1,2} \and Vivek Kulkarni\inst{1} \and Vasileios Theodorou\inst{3} \and Anton Matsiuk\inst{4}\and Simon Kuenzer\inst{4} \and Nikolaos E. Petroulakis\inst{5} \and Konstantinos Fysarakis\inst{5}}
	%
	%
	%
	\institute{Siemens AG, Munich, Germany\\
		\and
		Technical University of Munich, Munich, Germany\\
		\and
		Intracom SA Telecom Solutions, Athens, Greece\\
		\and
		NEC Europe Ltd., NEC Laboratories Europe, Heidelberg, Germany\\
		\and
		Foundation for Research and Technology-Hellas, Greece
	}
	\vspace{-15mm}
	\maketitle              

	\begin{abstract}
		\vspace{-7mm}
		VirtuWind proposes the application of Software Defined Networking (SDN) and Network Functions Virtualization (NFV) in critical infrastructure networks. We aim at introducing network programmability, reconfigurability and multi-tenant capability both inside isolated and inter-connected industrial networks. Henceforth, we present the design of the VirtuWind architecture that addresses the requirements of industrial communications: granular Quality of Service (QoS) guarantees, system modularity and secure and isolated per-tenant network access. We present the functional components of our architecture and provide an overview of the appropriate realization mechanisms. Finally, we map two exemplary industrial system use-cases to the designed architecture to showcase its applicability in an exemplary industrial wind park network.
	\end{abstract}

	\vspace{-9mm}
	\section{Introduction \& Background}
	SDN and NFV promise the programmable connectivity and rapid service provisioning\cite{mwc2017}. However, in their current state, a number of modifications to the state-of-art SDN-/NFV-architectures are required to accommodate the requirements of critical infrastructure providers. VirtuWind aims to fill this gap by defining a unified SDN- and NFV-architecture that provides for QoS-constrained end-to-end (E2E) connectivity in intra- and inter-domain connectivity scenarios.

	A typical power control system comprises a collection of various monitoring and control components connected to a remote operator's grid control system. For example, in wind parks Supervisory Control and Data Acquisition (SCADA) is the main monitoring and management component deployed locally on-site. It is utilized for data collection and analytics, as well as for the remote configuration of setpoints of the turbine-internal controllers. The SCADA server is a sub-component responsible for controlling the power output of multiple different wind turbines, and adaptation of the total power output to the requirements received from the grid operator. Based on our traces from an operational wind park, a full cycle of a correct SCADA control loop execution (i.e. the collection of sensor measurements and SCADA's response) has a periodicity of $100-200ms$, from which we derive the uni-directional E2E delay requirement of $<\sim30ms$. Network availability requirements of a wind park correlate with the application availability requirement. The fail-over time of $\sim50ms$ \cite{alcalucent}, as well as the requirement of $99,99\%$ availability (equaling 50min downtime p.a. \cite{5genergy}) impose an important failure resilience task for both SDN control-and data-planes.

\begin{figure}[htb]
		\centering
		\makebox[\textwidth][c]{\includegraphics[width=1.0\textwidth]{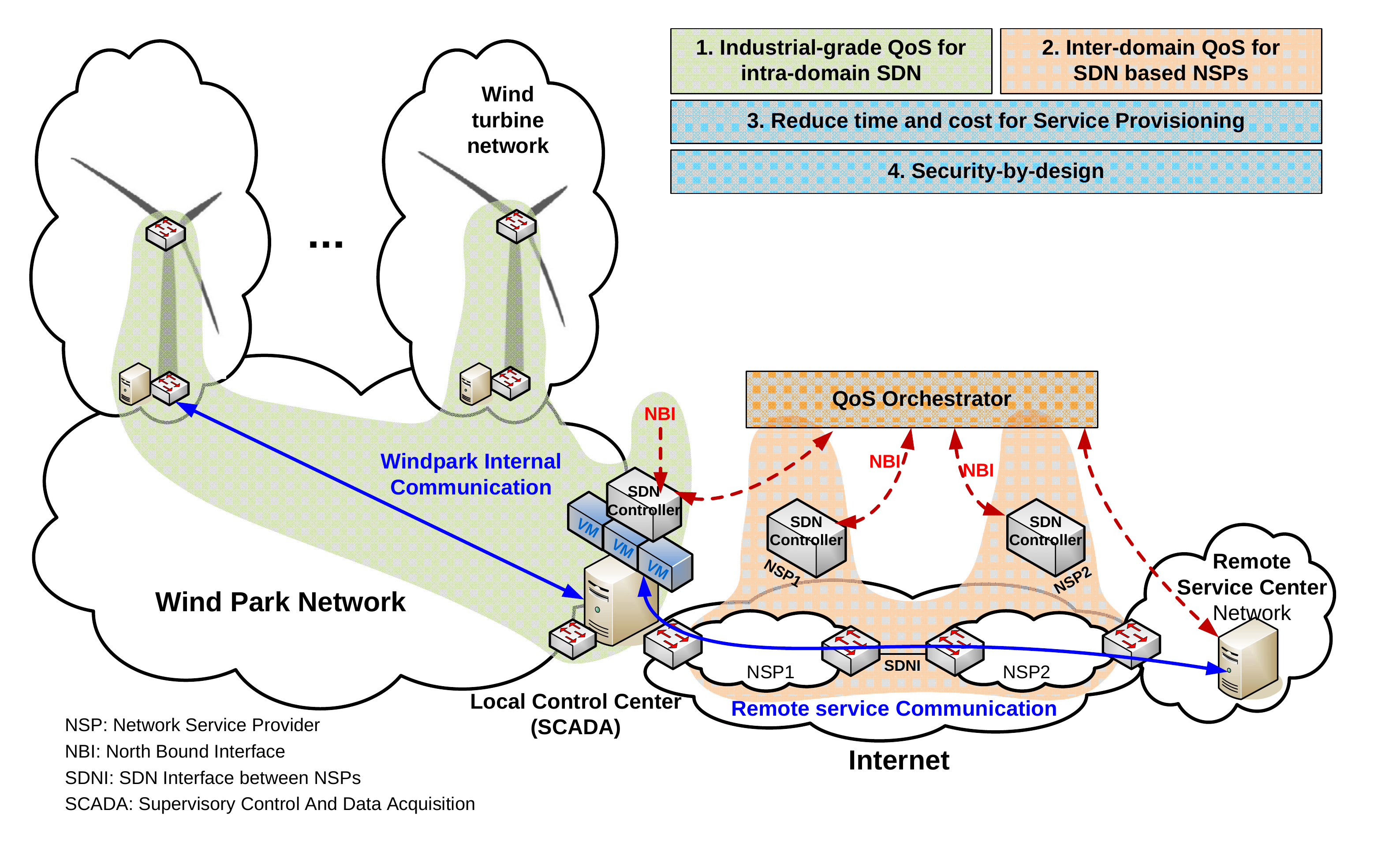}}
		\caption{VirtuWind Architecture Objectives}
		\label{fig:objectives}
	\end{figure}

	The VirtuWind architecture aims to fulfill the four key objectives depicted in Fig. \ref{fig:objectives}, using a combination of SDN- and NFV-technologies in a multi-operator ecosystem. We define three deployment steps necessary to enable the appropriate solution. The corresponding architecture is then derived in Sec. \ref{archsection}.

	\textbf{Step 1 - System Bootstrapping}: The deployed wind turbines are often dislocated from the centralized SCADA, thus deployment of an \emph{in-band} (shared) control network represents a strict requirement, opposite to the \emph{out-of-band} (exclusive) SDN controller-switch links often encountered in the data-center SDN deployments. Second, static appliances such as the data historians that collect and store large amounts of wind turbine monitoring data, should be placed so to optimize the resource sharing (i.e. the storage and compute resources). In the current non-virtualized deployments, the data historian components are deployed on dedicated industrial PCs inside the wind turbines. Thus, large cost savings are achievable by the centralization (and appropriate high-availability mechanisms in place) of such software appliances on dedicated "micro-cloud" nodes. Third, security components, such as firewalls, must be deployed for securing the external access to the industrial intra-domain network. 

	\textbf{Step 2 -  Enabling Intra-Domain Connectivity}: Traditionally, during system updates/upgrades the risk of impacting the SCADA control process is high, but for the regulatory reasons updates need to be performed regularly within a given time-frame, e.g. adapting the SCADA process services invoked by application or new service updates. The specific requirements in an industrial architecture come from e.g. an exemplary device upgrade workflow: 1) Maintenance tenant initializes a firmware update process using the SCADA interface; 2) The request is handled by the authentication entity in SCADA and the user is allowed or denied the service access. 3) If the user's request is accepted, the part of SCADA responsible for interactions with the SDN Controller triggers an event and request network "slice" with fixed time schedule and QoS support. 

	\textbf{Step 3 - Enabling Inter-Domain Connectivity}: The third deployment step enables the management of multiple wind park sites from remote locations, e.g. by a third party grid operator. To ensure a successful remote management, an inter-domain QoS enabled E2E connectivity is required. We foresee a centralized QoS approach, where an inter-domain coupling of SDN controllers and a centralized QoS orchestrator enables the E2E connectivity. Our approach involves four phases: 1. \emph{Domain Registration}; 2. \emph{Announcement of network path segments}; 3. \emph{Centralized E2E path computation} and 4. \emph{Path establishment}. 


	%
	\vspace{-4mm}
	\section{The VirtuWind Architecture}
	\label{archsection}
	\begin{figure}[htb]
		\centering
		\makebox[\textwidth]{\includegraphics[width=1\textwidth]{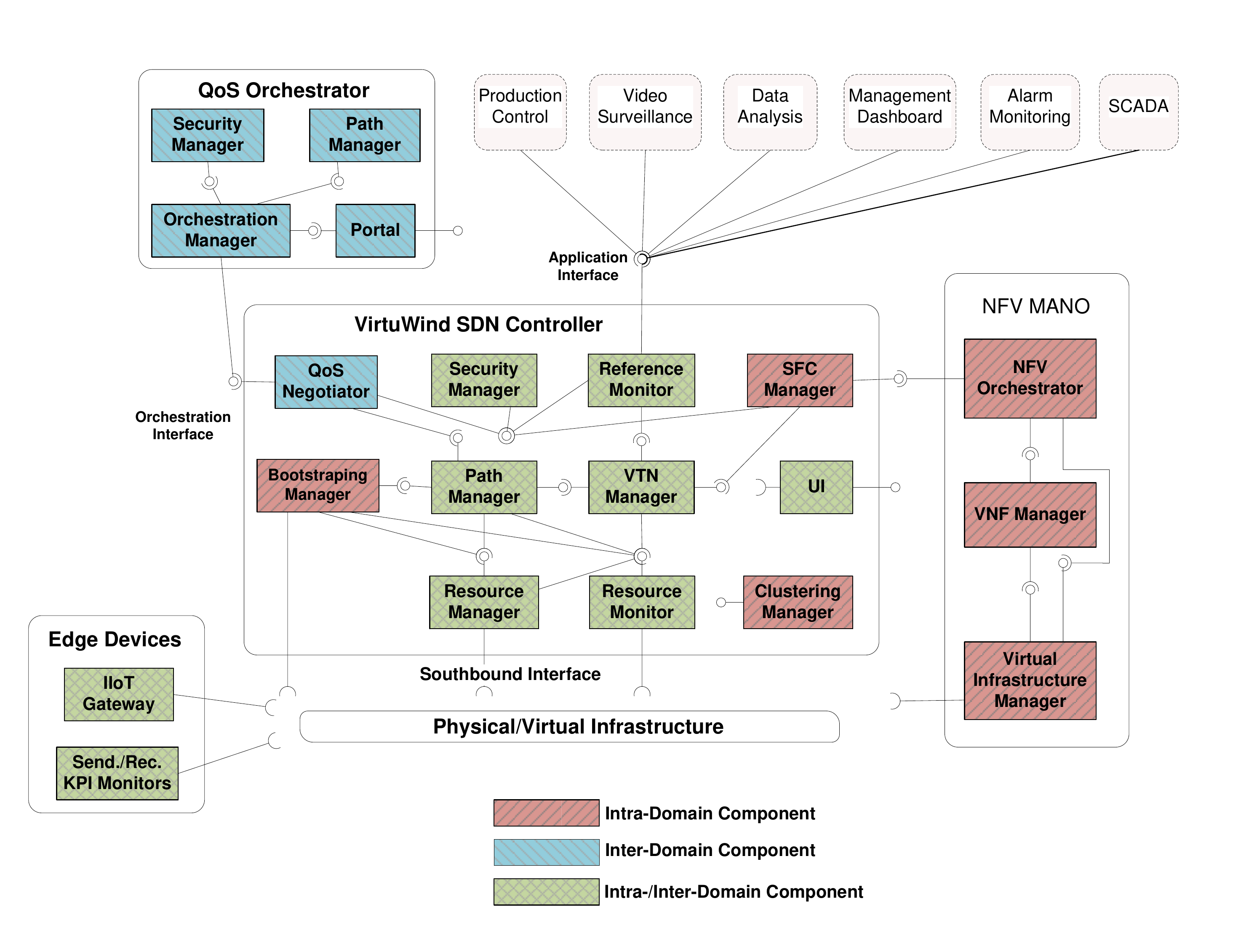}}
		\vspace{-5mm}
		\caption{The combined VirtuWind intra- and inter-domain system architecture. }
		\label{fig:architecture}
	\end{figure}
	\vspace{-4mm}
	VirtuWind envisions a layered architecture leveraging the control and programmability offered by the SDN paradigm and exploiting the flexibility of NFV. Each authorized application or tenant that needs connectivity requests an appropriate Virtual Tenant Network (VTN), and later issues communication service flow requests to the system. The flow requests are a combination of different connectivity and QoS service requirements, ranging from E2E-delay and bandwidth requirements, to different path protection schemes (e.g. duplication or fast-failover). The isolation of tenants is administered using the VTN northbound interface of the SDN Controller, while the service requirements define a communication service interface as per tenant's intent specification. Each application service is mapped to a unique VTN and a network tenant.
	\\
	Fig. \ref{fig:architecture} depicts the key architecture blocks of the VirtuWind architecture: 1) \emph{Business Applications} that interact with the underlying network and impose new service requirements at the centralized controller; 2) \emph{SDN Controller} and \emph{QoS Orchestrator}, which receive application requests, execute the centralized decision-making and configure the infrastructure; 3) \emph{NFV Management and Orchestration (MANO)} which orchestrates virtual network functions and service function chains of the industrial network; 4) \emph{Edge Devices} that allow for stretching of VTNs down to the last hop, as well as intent monitoring.

	In the remainder of this section, we give an overview of the components depicted in Fig. \ref{fig:architecture} and discuss their internal-workings. In Sec. \ref{workflowsection} we then outline the mapping of two exemplary processes on top of the introduced components.
\setlength{\parindent}{2ex}
	\vspace{-4mm}
	\subsection{Bootstrapping Manager}
	Industrial SDN networks require a highly-available, \emph{in-band} control plane \cite{vestin2015resilient}. By means of an automated network bootstrapping procedure, VirtuWind guarantees a robust and resilient control plane configuration during the network runtime. The robustness to controller failures is ensured by bootstrapping a multi-controller state replication design. To handle the data plane failure effects on the control plane flows, we leverage redundant control flow embedding. While recent works propose slower, restoration-based techniques in industrial scenarios \cite{vestin2015resilient}, we use 1+1 protection \cite{kirrmann2009hsr} by duplicating controller-to-controller and controller-to-switch TCP-based flows on maximally disjoint paths, thereby ensuring zero packet loss for control flows, at the expense of doubled bandwidth requirements per control flow connection. Since these typically have low bandwidth requirements, we do not consider it a crucial drawback in our approach. 
	\vspace{-4mm}
	\subsection{Reference Monitor (RFM)}
	RFM \cite{refmon} is a component that interacts with the northbound applications via the North-bound Interface (NBI). 
	NBI is designed in line with the principles of “Intent APIs” \cite{onfintent}, hiding complexity of the underlying infrastructure.

	\textbf{Intra- and Inter-domain}: RFM is involved in authentication of the applications and authorization of their service requests. During application authentication phase, RFM acts as a proxy between the application and Security Manager (SM). It receives initial credentials from the application and passes them to the SM. In case of a successful authentication, all subsequent application requests are evaluated against the authentication credentials approved by the SM. For authorization purposes all communication requirements contained in the service request are checked against pre-configured application access profiles. 

	\textbf{Inter-domain}: In inter-domain service setups, RFM additionally parses inter-domain requests 
	and, if communication endpoints are located in different domains, splits application requests into intra- and inter-domain parts. The intra-domain part of an initial application request is forwarded to VTN Manager (VTNM) in which RFM specifies the VTN’s gateway as the endpoint in the intra-domain part. The inter-domain part of the request is forwarded to QoS Negotiator and contains information about endpoint domains and service request requirements including QoS metrics. If QoS Negotiator receives a reply by the QoS Orchestrator that the inter-domain part requirements cannot be met, based on announced resources of all registered NSP domains, the request is declined. 
	\vspace{-4mm}
	\subsection{VTN Manager (VTNM)}
	\label{vtnmanager}
	
	\indent	

	\textbf{Intra-domain}: VTNM represents an extension of an existing framework\footnote{OpenDaylight's VTN project: \url{https://wiki.opendaylight.org/view/VTN:Main}} that realizes network slicing and exposes a set of corresponding APIs. By means of the VTN APIs one can create isolated virtualized network slices (VTNs) and provide L2/L3/L4 network forwarding functionality for such slices via virtual API primitives. VTNM maps these virtualized slices into physical infrastructure and enables forwarding via flow rules, at the same time ensuring slice isolation. The VTNM interacts with the Path Manager (PMG) to request for path computation of best-effort and mission-critical paths. Additionally, VTNM provides a VTN specification interface for the IIoT Gateway (ref. Sec. \ref{iiotgw}).

	\textbf{Inter-domain}: A specific virtual "gateway" interface is defined in every VTN, that is mapped to a physical interface of the domain’s border gateway, which serves as an exit point of the particular VTN in inter-domain.
	\vspace{-4mm}
	\subsection{Path Manager (PMG)}
	\label{pathman}
	\indent

	\textbf{Intra-domain}: For the guaranteed industrial QoS, i.e. the bandwidth provisioning, flow isolation and worst-case delay estimation, VirtuWind proposes using network calculus, a deterministic mathematical modeling framework for communication networks. Instead of basing its routing decision on a reactive control loop of network observations, VirtuWind's PMG provides mechanisms for admission control of new flows. By maintaining an accurate model of the network state and service embeddings in the control plane \cite{guck2014achieving}, PMG ensures per-flow isolation and worst-case guarantees at all times.

	\textbf{Inter-domain}: PMG handles all path configurations that need to take place on the network service provider's (NSP) network slice that is available to the VirtuWind architecture. In cross-NSP domains, traffic can flow through transit domains, for which QoS characteristics fluctuate over time in an unpredicted fashion. Thus, it is necessary for the NSP to continuously monitor the network performance and to be capable of applying corresponding actions once abnormal deviations are diagnosed. To this end, PMG maintains a detailed internal view of the network resources, and each NSP exposes to the QoS Orchestrator an aggregated view of the available path segments available within its domain. 

	\vspace{-4mm}
	\subsection{Resource Manager (RMG) and Resource Monitor (RMT)}
	\label{resmanager}
	
	\indent 
	
	\textbf{Intra-domain RMG}: RMG is responsible for configuration management and network control tasks, i.e. embedding of L2/L3 OpenFlow flow rules into the network. RMG provides for embedding of: i) real-time flows which require dedicated per-queue flow assignments; ii) best effort flows, without queue considerations; and iii) the meter structures for policing purposes. The generated flow rules and meter structures are persisted in the distributed data-store.
 
	\textbf{Intra-domain RMT}: RMT is a utility component that monitors and exposes the network state information. It provides for functionality to fetch the topology, as well as the features of the forwarding devices, providing input for PMG's routing and flow-queue mapping decisions. Furthermore, it allows for real-time monitoring of KPIs related to served intents.

	\textbf{Inter-domain RMG}: 
	The main difference compared to the intra-domain RMG functionality is in the use of match filters (e.g. MPLS labeling and VTN tagging), in order to enhance the scalability over large infrastructures. 

	\textbf{Inter-domain RMT}: For operator networks, due to scalability concerns, RMT performs real-time monitoring using probing and OpenFlow statistics.


	\vspace{-4mm}
	\subsection{Security Manager (SM)}
	\label{secman}
	\indent

	\textbf{Intra-domain}: SM realizes the authentication and accounting services to the rest of the SDN Controller as well as the users and applications that interact with the controller. With respect to authentication, the SM exposes interfaces for the administration of local SDN Controller accounts. Additional APIs are exposed for applications to present their credentials. If these credentials prove valid, the SM can issue an authentication token to the requesting party. The token can then be presented to the RFM when attempting to interact with the SDN Controller. The RFM is responsible for transferring these tokens to the SM internally for validation, so the former can then proceed to evaluate the request (i.e. if it is allowed based on the active policies). In the case of distributed authentication, the SM is responsible for presenting the tokens to the server for validation. 

	\textbf{Inter-domain}: 
	To authenticate all entities in an inter-domain scenario, two different approaches can be applied: the direct authentication on the controllers and the federated authentication via a trusted third party, which acts as an identity provider. In the former case, QoS Orchestrator that requests path segment offers, must have credentials on each controller to be authenticated locally. This creates the complexity of cross controllers entity credential data synchronization, so the SM module is extended to add the latter approach. In the case of distributed authentication setups, whereby the QoS Orchestrator 
	has an account created at another server 
	, the Token Bearer\cite{jones2012} authentication is used. QoS Orchestrator can be authenticated by requesting an authentication token by the OpenID Connect server. When a request arrives by the QoS Orchestrator to the QoS Negotiator of the controller, it passes the authentication token to the SM to verify it. The SM then validates the provided token at the identity provider.

	\vspace{-4mm}
	\subsection{Clustering Manager (CMG)}
	\label{clustering}
	The issue of the controller's single point of failure is resolved by means of state replication and fail-over to a pre-configured backup controller on failure. 

	\emph{Centralized controller state registry}: The CMG handles the controller relationships per-data state in the VirtuWind's distributed data-store. VirtuWind's controller state as well as the up-to-date network information is collected in a single registry shard that is replicated across the multiple controller instances. 

	\emph{Strong (SC) and adaptive consistency (AC) primitives for update ordering}: Components that have stringent requirements on the data state staleness, such as the Path Manager which makes critical routing and resource reservation decisions, may require serialized updates. Serialization ensures no data-store updates are applied without having first observed the previous history of the updates made to that state. Such components make use of the controller state distribution based on \emph{SC primitives} (e.g. on RAFT \cite{raft} consensus). Components that tolerate a certain degree of inconsistency may rely on \emph{AC primitives}. Our AC framework enables eventually consistent state synchronization with staleness bounds \cite{sakic2017towards}. The staleness bounds are realized by limiting the amount of de-synchronization in between controller replicas.

	\vspace{-4mm}
	\subsection{Service Function Chaining Manager (SFCM)}
	\label{sfcmanager}
	In industrial networks, Virtual Network Functions (VNFs) such as Firewall, IDS, DPI, and honeypot are pertinent to ensuring secure multi-tenant operation. SFCM is able to handle VNF chaining. It  invokes the VTNM in order to register external ports of the SDN transport network and to declare and associate service instances to those external ports. It exposes an interface to fetch information on and modify the existing service chains, the VTN-to-SFC mappings, as well as the service instances of the VNFs. Having the SFCM as a separate from MANO offers the advantages of having one interface to business applications, and the application does not need to be aware of the underlying SFC.

	\vspace{-4mm}
	\subsection{NFV Management and Orchestration (MANO)}
	\label{nfvo}
	VirtuWind does not require an implementation of a fully-fledged NFVO in the broad sense, as an inter-domain orchestration of Virtual Network Functions (VNFs) still lacks an appropriate use-case in the industrial domain. Nevertheless, we assume an implementation of an NFVO element (i.e. the OpenStack's HEAT-API) to stay compatible to the industry standard when VNF and SFC request specification is concerned. The VNFs deployed in industrial networks typically require static, long-lived configurations, but may need appropriate pre-configurations for correct security service function chaining. Provisioning of VNF requests is initiated by the user at the NFVO component and the requests are forwarded to VIM in order to request the deployment of the VNFs. When a new VNF instantiation request is received at the HEAT components, it interacts with the VIM to configure the infrastructure in order to provision the VNF. VIM guarantees the management and the allocation of the necessary virtual resources for the VNF deployment. Similar to data-center networks, the VIM is responsible for allocating the necessary resources in an industrial "private cloud". 
	\subsection{QoS Orchestrator (QOR)}
	\label{qosorchestrator}
	QoS Orchestrator (QOR) is responsible for setting up a QoS-enabled end-to-end connectivity service via multiple network operator domains. There are four phases in the lifecycle of a QoS-enabled end-to-end connectivity service: 1) \emph{Registration}: An NSP registers its domain to the QOR.
	2) \emph{Path Segment Announcement}: The registered NSP advertises its available resources.
	3) \emph{Path Segment Instantiation}: The QOR instructs the NSP to assign resources for a path.
	4) \emph{Monitoring}: The NSP periodically sends performance statistics to QOR to verify that agreed constraints are met for allocated flows.

	\textbf{Orchestration Manager}: Coordinator of all QOR activities, handling all communication with the SDN Controllers via encrypted REST APIs.

	\textbf{Path Manager}: Implements all path calculation logic for the establishment of inter-domain paths. Using end-point addresses as input, as well as the advertised path segment offers and corresponding QoS values, Path Manager finds zero or more ``best'' end-to-end paths (consisting of multiple NSP path segment offers) that satisfy specified QoS properties. 

	\textbf{Security Manager}: Implements the authentication and authorization to secure the communication between the QOR and the SDN Controller. 

	\textbf{Portal}: Front-end for the QOR administrator to trigger QOR functions manually and to visualize status information at runtime.
	\vspace{-4mm}
	\subsection{QoS Negotiator (QON)}
	\label{qosnegotiator} 
	QoS Negotiator (QON) is responsible for the communication between the SDN Controller and the QOR and the translation of QOR's requests to domain-specific actions. After registration, the QON receives path segment offer announcement and instantiation requests from the QOR and upon authentication and authorization via Security Manager, it replies to the QOR and propagates its requests into network actions. The information revealed to the QOR is high level enough so as not to expose sensitive internal details about the intra-domain configuration and characteristics
    . In this respect, path segment announcements only expose information about available path IDs, their ingress and egress border devices and their expected QoS characteristics. In addition, the QON periodically receives monitoring messages and notification events from the Path Manager, processes and filters them and sends to the QOR relevant information
    . This information comprises critical updates about the topology, e.g., path segments becoming unavailable or potential threats to maintaining the advertised QoS.
	\vspace{-4mm}
	\subsection{Industrial IoT Gateway (IIoTG)}
	\label{iiotgw}
	IIoTGs are physical devices that seamlessly integrate sensor devices to the VirtuWind network. Each IIoTG maps the sensors to VTNs and takes care of the isolation of the tenants when interacting with the sensors. An IIoTG is connected on one both to the industrial SDN network and at least one sensor network. It is then able to distinguish data flows from sensors and to forward them appropriately to assigned tenant networks. This design enables integrating sensor networks to the VirtuWind architecture that are not SDN-programmable and cannot implement any isolation (e.g., LoRa devices). A software SDN switch operating on the IIoTG’s hypervisor seamlessly integrates VNFs into the VirtuWind network. A VNF is instantiated for each sensor(s)--tenant pair and represents a set or a single sensor within a tenant. NFV is approached by exploiting the idea of \emph{unikernels}, the purpose-built operating systems that run a single targeted application. Unikernels utilize remove unnecessary kernel components in order to reduce the required resource usage \cite{minicache2017}. This enables running a high number of virtual instances on a single resource-constrained IIoTG\cite{minicache2017}. 

	\vspace{-4mm}
	\section{Exemplary System Workflows}
	\label{workflowsection}
	This section portrays the suitability of VirtuWind architecture to common use cases of isolated virtual tenant network addition, as well as the QoS-constrained network service embedding that spans multiple VirtuWind domains.

	\begin{figure}[htb]
\vspace{-7mm}
  \centering
      \makebox[\textwidth][c]{\includegraphics[width=0.9\textwidth]{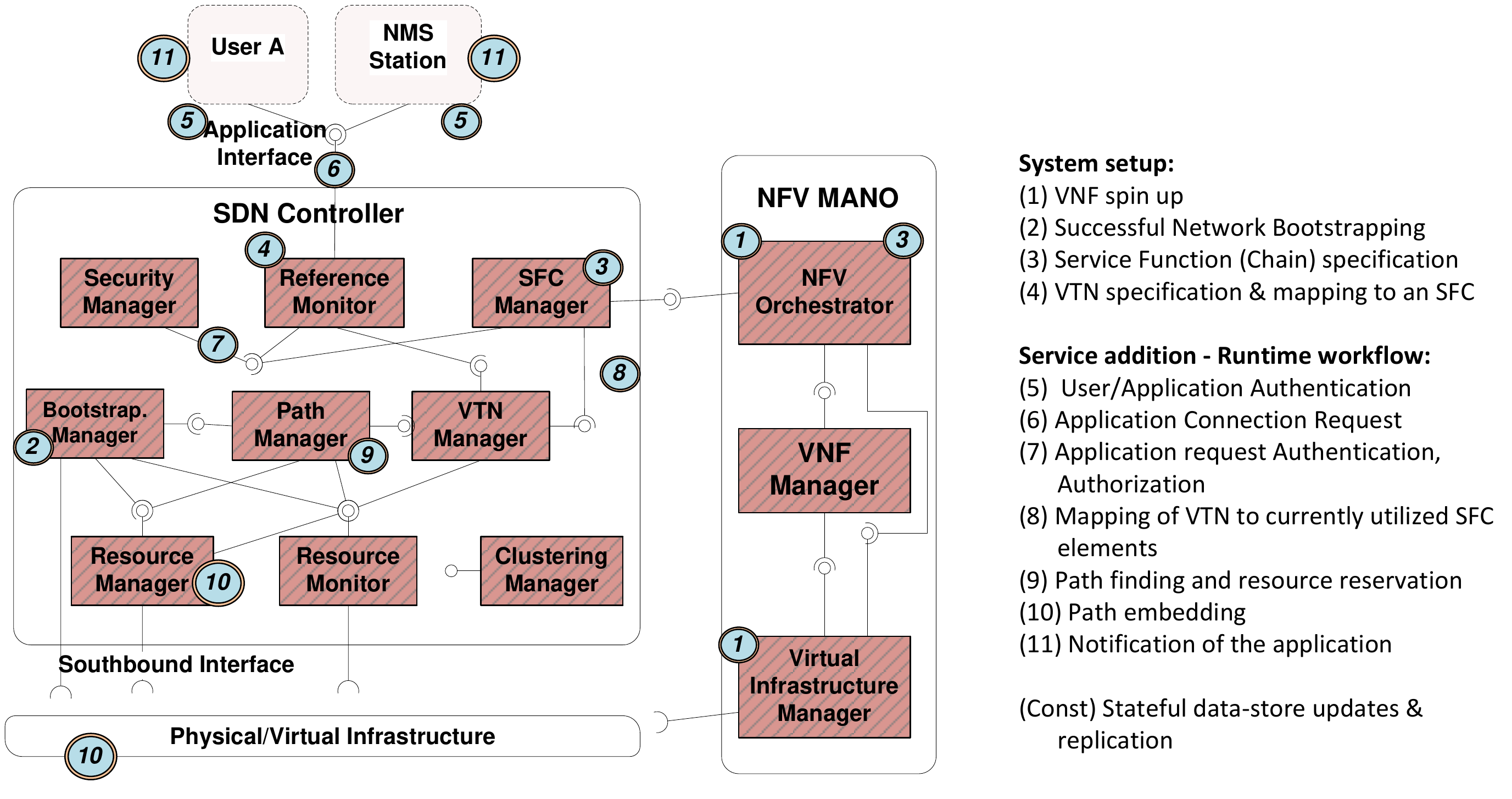}}
  \vspace{-6mm}
  \caption{An exemplary intra-domain VirtuWind system workflow. }
  \label{fig:intra-workflow}
\end{figure}
	\begin{figure}[htb]
\vspace{-12mm}
  \centering
  \centering
      \makebox[\textwidth][c]{\includegraphics[width=0.9\textwidth]{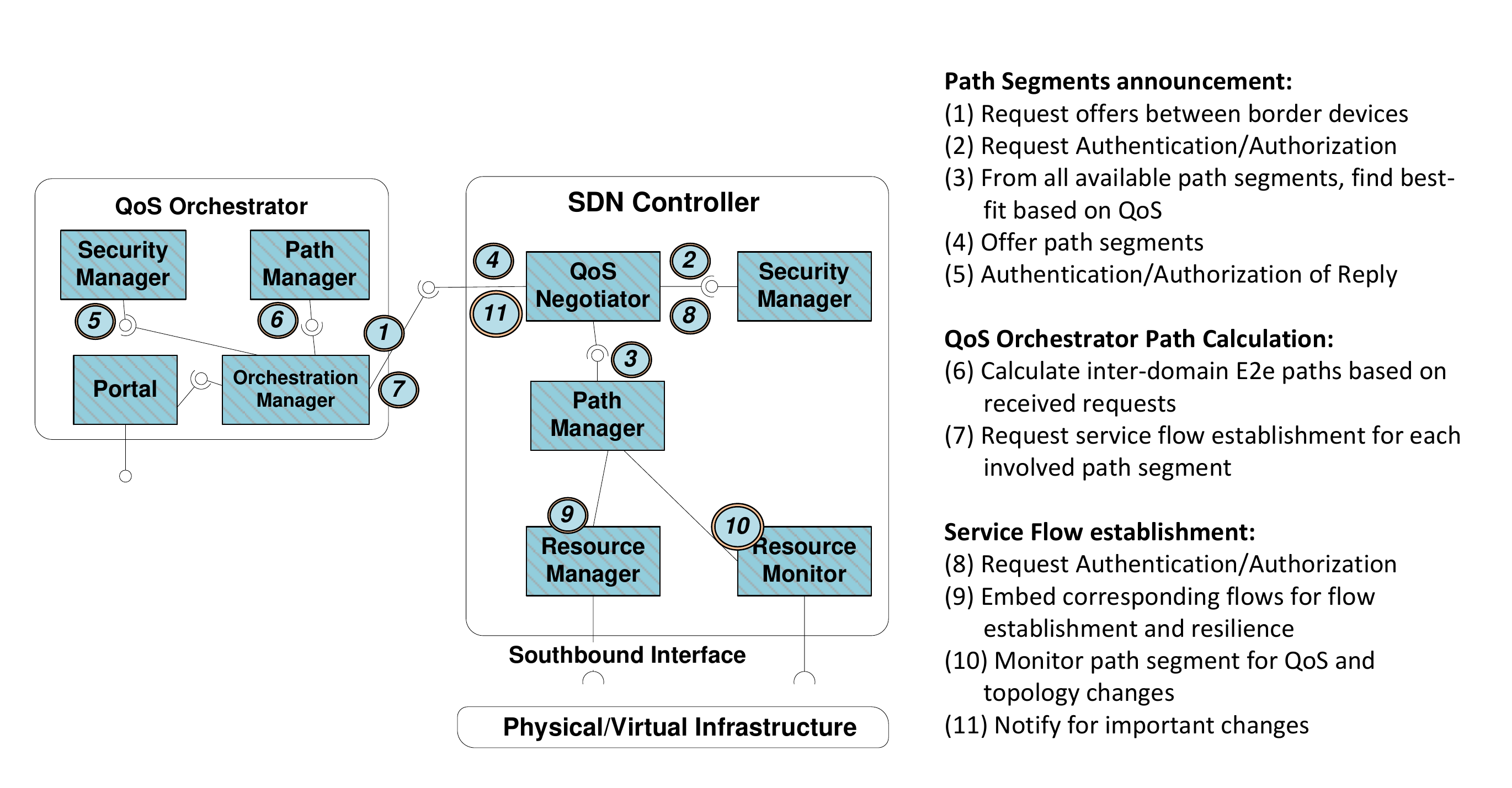}}
  \caption{An exemplary inter-domain VirtuWind system workflow. }
  \label{fig:inter-workflow}
\end{figure}

	Fig. \ref{fig:intra-workflow} depicts the scenario in which a new critical infrastructure service addition, that considers a set of QoS requirements specific to the \emph{User A}'s applications, is embedded into the SDN network. In addition to User A, an NBI-aware application (e.g. the depicted Network Management Station), may schedule and enforce a QoS-constrained intent at any point at network runtime.

	Fig. \ref{fig:inter-workflow} shows the inter-domain end-to-end path establishment scenario. It requires the interaction between the QoS Orchestrator and all the SDN Controllers of the NSP domains involved (we depict only one of those SDN Controllers) and it consists of the relevant service flow establishment for each involved NSP domain, combined with the path embedding for the endpoint industrial domain.

	\vspace{-4mm}
	\section{Conclusion} 
	\label{conclusion}

	This paper presents the VirtuWind architecture, that addresses the complex real-world requirements of an industrial network. The representative wind park control use case is briefly described and the design of an SDN- and NFV-architecture which considers industrial intra- and inter-domain requirements is illustrated in detail. We elaborate the mechanisms of individual components of the architecture and present their mapping to the architecture components. 
	\\
	\\
	\emph{\textbf{Acknowledgement}: This work has received funding from the EU's H2020 research and innovation programme under grant agreement No. 671648 VirtuWind.}


	%
	%
	\vspace{-4mm}
	\bibliographystyle{splncs} 
	\bibliography{bibl}
\end{document}